\documentclass[letterpaper,twocolumn,english,aps,prl,superscriptaddress,floatfix]{revtex4}
\usepackage{amsmath}
\usepackage{graphicx}

\renewcommand{\[}{\begin{equation}}

\renewcommand{\]}{\end{equation}}

\usepackage{babel}

\usepackage{babel}

\usepackage{babel}

\usepackage{babel}

\usepackage{babel}

\usepackage{babel}

\begin{document}
\global\long\def\avg#1{\langle#1\rangle}

\global\long\def\p{\prime}

\global\long\def\dg{\dagger}

\global\long\def\ket#1{|#1\rangle}

\global\long\def\bra#1{\langle#1|}

\global\long\def\proj#1#2{|#1\rangle\langle#2|}

\global\long\def\inner#1#2{\langle#1|#2\rangle}

\global\long\def\tr{\mathrm{tr}}

\global\long\def\pd#1#2{\frac{\partial#1}{\partial#2}}

\global\long\def\spd#1#2{\frac{\partial^{2} #1}{\partial#2^{2}}}

\global\long\def\der#1#2{\frac{d #1}{d #2}}

\global\long\def\im{\imath}

\renewcommand{\onlinecite}[1]{\cite{#1}}

\title{Quantized ionic conductance in nanopores}

\author{Michael Zwolak}

\affiliation{Theoretical Division, MS-B213, Los Alamos National Laboratory, Los
Alamos, NM 87545}

\author{Johan Lagerqvist}

\affiliation{Department of Physics, University of California, San Diego, La Jolla,
CA 92093}

\author{Massimiliano Di Ventra}

\affiliation{Department of Physics, University of California, San Diego, La Jolla,
CA 92093}

\date{\today{}}
\begin{abstract}
Ionic transport in nanopores is a fundamentally and technologically
important problem in view of its occurrence in biological processes
and its impact on novel DNA sequencing applications. Using microscopic
calculations, here we show that ion transport may exhibit strong nonlinearities
as a function of the pore radius reminiscent of the conductance quantization
steps as a function of the transverse cross section of quantum point
contacts. In the present case, however, conductance steps originate
from the break up of the hydration layers that form around ions in
aqueous solution. Once in the pore, the water molecules form wavelike
structures due to multiple scattering at the surface of the pore walls
and interference with the radial waves around the ion. We discuss
these effects as well as the conditions under which the step-like
features in the ionic conductance should be experimentally observable. 
\end{abstract}
\maketitle
Over the last decade there have been tremendous advances in both the
fabrication of nanopores and their use for molecular recognition and
nucleic acid analysis \citep{Zwolak08-1}. Experimental characterization
of molecules has mostly relied on measuring changes in the ionic current
through the pore \citep{Kasianowicz1996-1,Akeson1999-1,Meller2000-1,Vercoutere2001-1},
but other ways to probe single molecules in nanopores may come from
embedding nanoscale sensors within the pore or nanochannels \citep{Zwolak2005-1,Lagerqvist06-1,Lagerqvist07-1,Lagerqvist07-2,Heng2005-1,Gracheva2006-1,Gracheva2006-2,Liang08-1}.
However, electronic fluctuations due to the dynamical ionic and aqueous
environment will affect the type of signals and noise these sensors
detect. Therefore, understanding the electrostatics of ions in water
at atomic length scales is crucial in our ability to design functional
single-molecule sensors and to interpret their output, and will also
provide new insight into the operation of the ubiquitous biological
ion channels.

Several studies have examined the electrostatics of ions in nanopores
using continuum models for the dielectric properties of water \citep{Zhang2005-1,Kamenev2006-1,Teber2005-1,Bonthuis06-1}.
Within a continuum model, the nanopore electrostatic environment is
essentially one-dimensional due to the large difference of dielectric
constants between water and the surrounding pore material \citep{Teber2005-1}.
Thus, according to these models, there is a large electrostatic energy
penalty to move an ion from the exterior of the pore to its interior
\citep{Zhang2005-1,Teber2005-1}, with small amounts of surface charge
able to drastically reduce this energy penalty \citep{Zhang2005-1}.
Although continuum models can highlight some generic features of ionic
currents and blockades, such as the effect of surface charges, they
miss important effects related to the microscopic physical structure
of water molecules around ions.

\begin{figure}
\begin{centering}
\includegraphics[width=7cm]{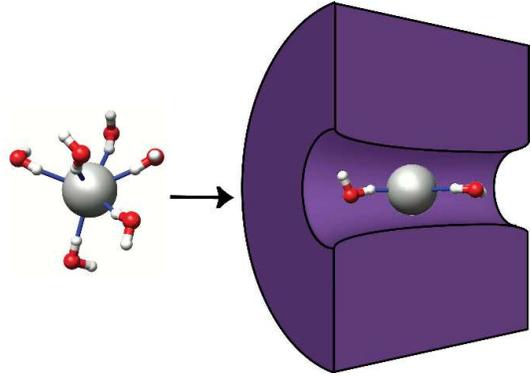} 
\par\end{centering}

\caption{Schematic of a single anion with its first hydration layer that moves
into a nanopore. Due to spatial constraints many water molecules need
to be stripped off for the ion to enter the pore, creating nonlinearities
in ionic conductance. \label{fig:hydration}}

\end{figure}

Here, we examine ionic transport from the point of view of these nanoscale
features (see schematic in Fig. \eqref{fig:hydration}). Using all-atom
molecular dynamics simulations, we calculate the structural and electrostatic
details of a single ion in an aqueous environment both in and out
of a cylindrical nanopore (see Methods for details). Ions in solution
create local structures in the surrounding water, known as \emph{hydration}
\emph{layers} \citep{Hille01-1}, and characterized by oscillations
in the water density. The length scales associated with these oscillations
and their decay are in the range $2-10\,\textrm{\AA}$, i.e., comparable
to the radial dimensions of some nanopores. In addition, we show that
when the ion is inside the pore, water molecules form wavelike features
due to both the interaction with the walls and the water molecules
in the partially broken hydration layers. Starting with these microscopic
considerations, we propose a model system based on the concept of
\emph{energy shells}, which incorporates the hydration layer structure
into the energetic barrier created by the pore. We predict that the
ionic conductance of a nanopore manifests step-like features as a
function of the pore radius similar to the quantized conductance of
quantum point contacts as a function of their effective cross section
(see, e.g., \cite{Diventra2008-1}). Finally, we examine the influence
of noise on ionic transport and discuss the experimental conditions
under which the above step-like features are most likely to be observed.

\begin{figure}
\begin{centering}
\includegraphics[width=8cm]{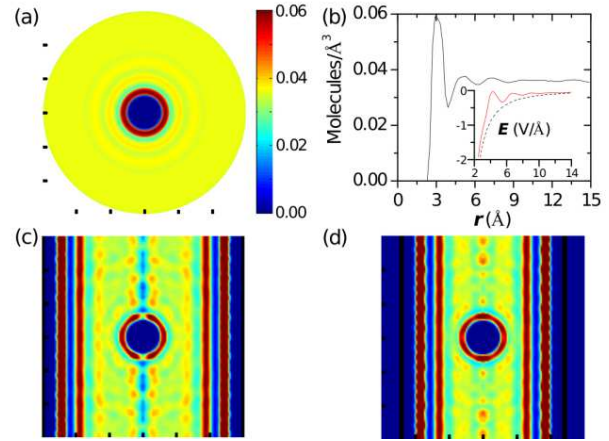} 
\par\end{centering}

\caption{Water density around a Cl anion both in bulk and inside cylindrical
nanopores of different radii. (a) Water density (using the oxygen
as the location of the water molecule) around a Cl anion in bulk water.
(b) Water density versus radial distance from a Cl anion. At distances
greater than $10\,\textrm{\AA}$ the water density is approximately
$0.036$ molecules$/\textrm{\AA}^{3}$, which is slightly more than
bulk water density of $0.033$ molecules$/\textrm{\AA}^{3}$ at room
temperature. Some edge effects are present around $15\,\textrm{\AA}$.
The distribution of water molecules shows oscillations corresponding
to the formation of hydration layers, which is also reflected in the
electric field. The inset shows the time averaged radial electric
field (versus the radial distance) from both the ion and water dipoles
(red line) and from just the ion (black dashed line). (c) Water density
around a Cl anion inside a $15\,\textrm{\AA}$ radius pore (the pore
walls are at the edge of the figure). There are wavelike features
that are due to interference patterns between oscillations reflecting
off the walls of the pore (that set an effective pore radius) and
those around the ion. However, the first two layers around the anion
are still present. (d) Water density around a Cl anion inside a $12\,\textrm{\AA}$
radius pore (the pore walls are indicated by vertical, thick black
lines). Here the effective radius is approximately $5\,\textrm{\AA}$,
thus the second hydration layer interacts with the effective pore
wall and it is partly broken. \label{fig:radialdensity}}

\end{figure}

To be specific, we focus our attention on an isolated chlorine ion
but similar considerations will apply to other types of ions and situations.
Immediately around the Cl anion, the water molecules orient themselves
so that a single hydrogen from each molecule points inward toward
the ion (see Fig. \eqref{fig:hydration}). The distribution of water
molecules forms into layers as shown in Fig. \eqref{fig:radialdensity}.
The inner most layer is very tightly bound, and subsequent layers
are spaced at $2.0-2.3\,\textrm{\AA}$. These findings are in good
agreement with neutron diffraction and X-ray absorption experiments
that measure the radial distribution of water \citep{Ohtaki1993-1},
which give a Cl-O peak at $\sim3.1\,\textrm{\AA}$, as well as molecular
dynamics simulations performed with different force field parameters
\citep{Yang05-1}. In addition, we can acquire further information
not directly accessible from experiments, such as the microscopic
electric field due to the ions and water (shown in the inset of Fig.
\eqref{fig:radialdensity}(b)). This field shows oscillations corresponding
to the hydration layers and is similar to a series of alternating
charged surfaces. A continuum picture, however, would not capture
these microscopic details that are responsible for the conductance
steps we predict.

If the anion is now placed in the nanopore, the hydration layers are
affected. For large pores, the hydration layer structure is identical
to that in bulk water. However, as the pore is taken to nanoscale
dimensions, eventually the pore walls (or effective pore wall, which
may be, e.g., a layer of tightly bound water molecules) will force
the hydration layer to be partially broken because it can not fit
within the pore. Figures \ref{fig:radialdensity}(c, d) show this
effect. Within a $15\,\textrm{\AA}$ radius pore, the first and second
hydration layers are still present. On the other hand, for a $12\,\textrm{\AA}$
pore radius, the first layer is still intact, but the second layer
is partially broken. Indeed, it is almost a spherical shell cutoff
by the effective pore wall which comes at about $5\,\textrm{\AA}$
from the axis of the pore. In order to gain physical insight, we propose
the following model system that captures this essential aspect of
the presence of hydration layers around the ion.

\emph{Model System -} The hydration layers stay partially intact except
for the portion of the hydration surface cut off by the pore walls,
i.e., for layers with radii greater than the effective pore radius.
In other words, only the portion of the layer along the axis of the
pore remains intact. We consider then a set of surfaces placed at
each hydration layer, $i$, at radius $R_{i}$. Each surface represents
the area where the water dipoles fluctuate, giving the time-averaged
dipole layers. Ignaczak et al. \citep{Ignaczak99-1} have found that
for small water clusters around a Cl anion, the internal energy of
each water is approximately linear in the number of water molecules
in the cluster. Since interactions (van der Waals and electrostatic)
with a low dielectric pore are small compared to water-ion and water-water
interactions \cite{Yang07-1}, this suggests writing the internal
energy contained in a partially intact hydration layer as

\begin{equation}
U_{i}=f_{i}U_{i}^{o}\label{eq:layerEnergy}\end{equation}
 where $f_{i}$ is the fraction of the layer present in the pore and
$U_{i}^{o}$ is the energy difference of that stored in the intact
water layer and the water in bulk.

In our model, the energy barrier is due to the stripping off of a
fraction of the layer, $f_{i}$, i.e., the fraction of a spherical
surface at $R_{i}$ that remains within the pore of effective radius
$R_{p}$. When the ion translocates along the pore axis, the surface
area that remains in layer $i$ is given by \begin{equation}
S_{i}=2\Theta\left(R_{i}-R_{p}\right)\int_{0}^{2\pi}d\phi\int_{0}^{\theta_{c}}d\theta R_{i}^{2}\sin\theta\end{equation}
 where $\Theta\left(x\right)$ is the step function and $\theta_{c}=\sin^{-1}R_{p}/R_{i}$.
When $R_{p}<R_{i}$, the fraction of the surface left is \begin{equation}
f_{i}\left(R_{p}\right)=1-\sqrt{1-\left(\frac{R_{p}}{R_{i}}\right)^{2}}.\end{equation}
 The internal energy barrier as function of pore radius is then given
by \begin{equation}
\Delta U(R_{p})=\sum_{i}(f_{i}(R_{p})-1)U_{i}^{o}\,,\label{eq:dE}\end{equation}
 where the summation is over the layers. The free energy change of
hydration is dominated by the contribution from the total internal
energy contained within the hydration layers, $\sum_{i}U_{i}^{o}$,
plus a bulk dielectric contribution from water outside the layers.
Previous calculations give the total internal energy change in the
range -3.5 to -3.7 eV \cite{Chandrasekhar84-1,Duan03-1,Ignaczak99-1}.
The form of the electric field shown in Fig. \ref{fig:radialdensity}(b)
suggests that the time-averaged microscopic distribution of waters
can be viewed as a set of spherical layers of alternating charge.
Thus, the energy stored within each layer should be well estimated
by using a Born solvation calculation, which, by construction, will
sum to the total internal energy contribution . The physical content
of this calculation is that of the solvation of embedded {}``quasiparticles'':
the total solvation energy is given by solvating the ion; the energy
of the first layer is the difference of solvating the ion and solvating
the ion plus the first layer, i.e., solvating a {}``quasiparticle''
consisting of the ion and a cluster of waters. With this picture in
mind, the energy within each layer is given by \[
U_{i}^{o}=\frac{e^{2}}{8\pi\epsilon_{0}}\left(1-\frac{1}{\kappa}\right)\left(\frac{1}{R_{i}^{O}}-\frac{1}{R_{i}^{I}}\right)\]
 where $\kappa$ is the dielectric constant of water and $R_{i}^{I\,(O)}$
are the inner (outer) radii demarcating the hydration layer, and we
have used the effective ion charge of $1e$, which is consistent with
the field recovering its bare ion value after each hydration layer
(see inset of Fig. \ref{fig:radialdensity}(b)). From the water density
oscillations shown in Fig. \ref{fig:radialdensity}, these radii are
2.0 and 3.9 $\textrm{Å}$ for the first layer; 3.9 and 6.2 $\textrm{Å}$
for the second; 6.2 and 8.5 $\textrm{Å}$ for the third. The total
solvation energy is then -3.6 eV and the layer energies are -1.7,
-0.7, and -0.3 eV for the first, second, and third layers, respectively.
Immediately, we can understand why the third layer is absent within
the pore: van der Waals interactions \cite{Yang07-1} and entropic
contributions from the water structure \cite{StokesBook} are of the
same magnitude.

\begin{figure}
\begin{centering}
\includegraphics[width=8cm]{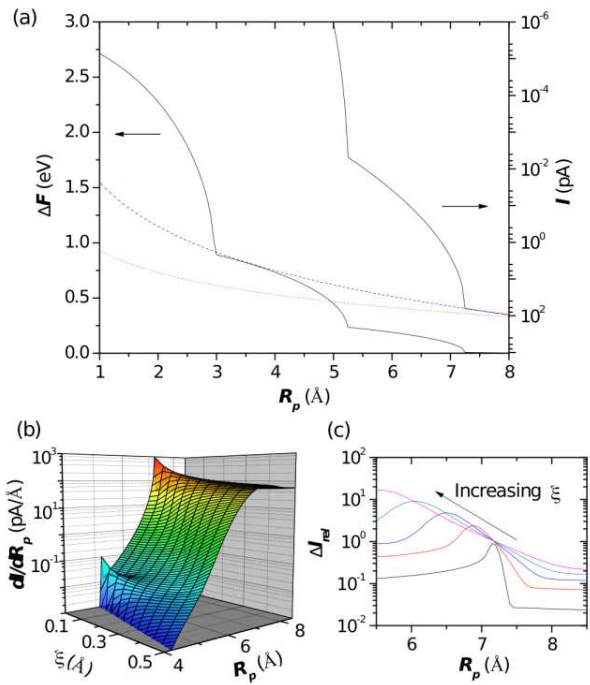} 
\par\end{centering}

\caption{(a) Free energy barrier to bring an ion into a nanopore and the ionic
current at room temperature as a function of the effective pore radius.
The dotted red line indicates the ionic current without a free energy
barrier and the dashed blue line is the current with just the entropic
barrier. The current is calculated at $120$ mV and $1$ M bulk salt
concentration assuming the ionic current is given by $I=2\pi R_{p}^{2}e\mu nE$,
where the density $n$ is proportional to a Boltzmann factor containing
the energy barrier, namely $n=n_{0}\exp(-\Delta F/kT)$, with $n_{0}$
the bulk salt concentration, $\mu$ the ion mobility, and $E$ the
strength of the electric field. A typical ion mobility is $\mu=2.28\times10^{-8}m^{2}/Vs$
(see, e.g., Refs. \citep{Deamer2002-1} and \citep{Ho2005-1}). (b)
$dI/dR_{p}$ as a function of pore radius and noise strength. For
zero noise, the step-like features in the ionic current give rise
to well defined peaks in the derivative of the current. However, as
the noise is increased, the peaks smooth out. This occurs beyond a
noise strength of $\xi=0.2-0.3\,\textrm{\AA}$ for the second and
third layers (see text for details). (c) Relative noise, $\Delta I_{rel}=\left(\left\langle I^{2}\right\rangle -\left\langle I\right\rangle ^{2}\right)^{1/2}/\left\langle I\right\rangle $,
versus pore radius for the outermost hydration layer and for $\xi=0.05,\,0.15,\,0.25,\,0.35,\,0.45\,\textrm{Å}$.
If the pore radius is tuned to the radius of a hydration layer, fluctuations
give rise to a peak in the relative noise, which gives an alternative
method to observe the effect of the hydration layer. \label{fig:energy}}

\end{figure}

In addition to the internal energy change, there is also an entropic
contribution to the free energy change that comes from removing a
single ion from solution and localizing it in the pore. Assuming an
ideal ionic solution, this entropic contribution is $\Delta S=k_{B}\ln\left(V_{p}n\right)$,
where $V_{p}$ is the volume of the pore, $n$ is the bulk ionic concentration,
and $k_{B}$ is the Boltzmann constant. The final free energy difference,
$\Delta F=\Delta U-T\Delta S$, is plotted in Fig. \eqref{fig:energy}.
This is the main result of this work: hydration layers hold energy
in a shell-like structure that causes step-like features in the energy
barrier to ionic transport. These step-like features generate corresponding
features in the ionic current shown in Fig. \eqref{fig:energy}.

For the remaining portion of this paper, we discuss the conditions
under which these features can be observed experimentally and address
how the assumptions we have made affect our conclusions. First, we
have assumed that only the single ion barrier is important, i.e.,
that there are no correlated transport processes. We can obtain an
indication of whether experiments are consistent with this assumption
by considering the ionic concentration. If the bulk salt concentration
is 1 M, a typical ion (counting both anions and cations) will be contained
in a volume with linear dimension 9.4 $\textrm{\AA}$. This is large
enough to almost house both the first and second hydration layers.
Since in the absence of significant surface charge on the pore walls,
one expects the ionic concentration within the pore to be lower, a
bulk concentration of 1 M is consistent with having ions translocate
through the pore independently. A 1 M concentration is likely to be
small enough to allow at least the effects of the first ionic layer
to be observed, i.e., a sudden drop in the current when the pore radius
is comparable to the first hydration layer. For arbitrary concentrations,
the ion-ion distance goes as $\sim9.4/n_{0}^{1/3}\,\textrm{\AA}$.
If the ion is to be contained in a linear dimension of $20\,\textrm{\AA}$
to allow all hydration layers to be present, then one would want a
concentration of about 0.1 M. Low concentrations are therefore key
to observe the outermost hydration layers. Furthermore, the ionic
concentration itself may be used to observe the step-like features.
Namely, at certain concentrations the anion-anion (cation-cation)
distance would be small enough so that the hydration layers should
break with increasing concentration, and a corresponding nonlinearity
would be observed.

Second, we have neglected energetic fluctuations due to other effects
present, such as thermal noise, ion-ion interaction, rough pore surfaces,
a distribution of ion paths through the pore, and the fact that hydration
layers are not exactly represented by their time-averaged oscillations
(therefore, some ions with intact hydration layers could go into pores
smaller than the radius $R_{i}$ would predict). In order to determine
how robust the step-like features are to these noise sources, we add
in our calculations Gaussian noise in the pore radius. In particular,
we define a noise strength parameter $\xi$ (i.e., the standard deviation
of the Gaussian noise) to describe both energetic and path fluctuations.
In Fig. \eqref{fig:energy}, we plot $dI/dR_{p}$ as a function of
$R_{p}$ and $\xi$. For no noise, $\xi=0\,\textrm{\AA}$, there are
well defined peaks in $dI/dR_{p}$. However, at a noise strength of
$\xi=0.2-0.3\,\textrm{\AA}$, the peaks from the second and third
layers are smoothed out. Given the width of the hydration layers as
seen in Fig. \eqref{fig:radialdensity}, the effects from the second
and third layers may not be observable when transport is due to ions
like Cl$^{-}$, or $ $K$^{+}$. However, other ions, such as the
divalent Mg or trivalent Al ions \cite{Ikeda07-1,Bylaska07-1}, as
well as some monovalent ions \cite{Yang07-1}, have more strongly
bound second and third layers. Together with using tunable properties,
such as temperature \cite{Bylaska07-1,Duan03-1}, this may allow observing
the step-like features in the current. In addition, the noise gives
an alternative way to observe the effect of the hydration layers.
As shown in Fig. \eqref{fig:energy}(c), the relative noise peaks
when the pore radius is near that of a hydration layer. This effect
can be understood in terms of a two-state picture: fluctuations that
decrease the pore radius energetically {}``close'' the pore, whereas
fluctuations that increase the pore radius {}``open'' the pore,
essentially creating an instability in the current. Such an effect
could also be induced artificially by sampling over a distribution
of pore radii centered around a layer radius. The other features of
the relative noise can also be understood within this picture.

Third, we have assumed ion translocation is along the axis of the
pore. In reality, the ions can move off center. There are different
factors which cause this. First, the hydration layers at any given
instant are not spherically symmetric. In fact, some ions have asymmetric
hydration layers where at a given time, for instance, the first layer
is offset to one side of the ion \cite{Ignaczak99-1}. However, this
fact cannot destroy the effect we predict: in these cases, the hydration
layer may be able to {}``squeeze'' into a smaller radius pore by
moving off center, but eventually it has to break in a highly nonlinear
way. Another cause, however, is something that our model predicts:
if the energy is contained in a spherical layer, then the ion actually
tends to move off center. This is a consequence of geometry: more
of the spherical surface can fit into the pore if the ion moves toward
the pore wall. However, the ion cannot move too far from the center,
otherwise some of the inner layers would have to be partly destroyed
with consequent energy penalty. Neither of these two causes would
change our predictions, although they may change some details, such
as at what pore radius the conductance jumps occur. Finally, a strong
affinity of the ion for the pore wall may occur, for instance, if
the pore were significantly charged. This effect may indeed reduce
the energy barrier for ionic transport considerably, making all but
the first hydration layer breakup undetectable in the ionic current.
The charging of the pore walls, which depends heavily on pore material
and treatment \cite{Chen2004-2}, as well as other factors, can again
be somewhat resolved by observing transport with different ion and
pore types.

Before we conclude, we briefly mention the other water patterns within
the pore. These patterns depend on pore size - different size pores
can give maxima and minima in the water density along the pore axis
\cite{Lynden96-1}. Further, we find that when the pore radius is
$10\,\textrm{\AA}$ (effective radius $\sim3\,\textrm{\AA}$) there
is a transition in the water structure of the pore \cite{Hummer01-1,Zangi03-1,Carrillo06-1}.
Carefully controlled ionic transport measurements may also be able
to detect these changes in addition to the nonlinear features due
to the hydration layers.

\emph{Conclusions -} We have examined ionic transport through nanopores
using both molecular dynamics and electrostatic calculations. A simple
form of the free energy barrier can capture an important effect of
hydration layers - namely that the layers produce step-like features
in the energy barrier for ions to enter a nanopore, and thus also
the ionic conductance, as a function of pore radius. This effect is
the classical counterpart of the electronic quantized conductance
one observes in quantum point contacts as a function of their cross
section. Like the quantum case, one needs sufficient experimental
control of the nanopore characteristics in order to observe this effect.
Irrespective, a drop in the current should be observable when the
first hydration layer is hit, similar to what is thought to occur
in the open and closed states of some biological pores \cite{Peter05-1,Miyazawa03-1,Beckstein04-1}.
However, under the right conditions, multiple current drops should
be observed. Further, the relative noise has a characteristic peak
when the pore radius is near a hydration layer radius, and this gives
an alternative way to observe the effect of hydration. Due to the
growing importance of ionic transport in DNA sequencing, this effect
may have consequences for future technologies. In addition, the operation
of biological ion channels has many factors associated with it, such
as favorable internal sites \cite{Gouaux05-1} that compensate for
the loss of hydration (favorable interactions can also exist in artificial
pores \cite{Fornasiero08-1,Nishizawa95-1}) or correlated ionic processes
\cite{Berne2001-1}. Thus, investigations with more well-controlled
synthetic pores can help shed light on the different factors at play
and we hope our work will motivate experiments to specifically examine
the role of dehydration.

\emph{Methods} - The molecular dynamics simulations are performed
with the NAMD2 package \cite{Kale1999-1} and the simulation details
are as in Refs. \cite{Lagerqvist06-1,Lagerqvist07-1,Lagerqvist07-2}.
The TIP3 \cite{Jorgensen1981-1} model is used for the description
of water, which is known to reproduce the dielectric properties of
bulk water \cite{Simonson1996-1}. For all simulations, a single chlorine
anion is taken as fixed since its thermal wavelength at room temperature
is $\sim0.2\,\textrm{\AA}$. The cylindrical, Si$_{3}$N$_{4}$ nanopore
surfaces carry a charge of $6-8\times10^{-4}$ e/atom for the $15\,\textrm{\AA}$
and $12\,\textrm{\AA}$ radius pores. The density plots for an ion
in a spherical droplet of water are calculated by averaging the last
900 out of a 1,000 snapshots (taken at 1 ps intervals) within spherical
shells of thickness $1\,\textrm{\AA}$. For an ion inside a pore,
the last 4000 out of 5000 snapshots (taken at 0.1 ps intervals) are
averaged within cylindrical shells of thickness $1\,\textrm{\AA}$
in both the radial and vertical directions. The electric field is
calculated by averaging the ion and dipole moment contributions over
the snapshots. 
\begin{acknowledgments}
This research is supported by the NIH-National Human Genome Research
Institute (JL and MD) and by the U.S. Department of Energy through
the LANL/LDRD Program (MZ). 

\bibliographystyle{apsrev}
\bibliography{DNAandPores}

\end{acknowledgments}

\end{document}